# Intracavity characterization of micro-comb generation in the single-soliton regime

**Pei-Hsun Wang[1,\*,†], Jose A. Jaramillo-Villegas[1,3,\*,†], Yi Xuan[1,2], Xiaoxiao Xue[1], Chengying Bao[1], Daniel E. Leaird[1], Minghao Qi[1,2], and Andrew M. Weiner[1,2]**

[1]*School of Electrical and Computer Engineering, Purdue University, 465 Northwestern Avenue, West Lafayette, IN 47907, USA*
[2]*Birck Nanotechnology Center, Purdue University, 1205 West State Street, West Lafayette, IN 47907, USA*
[3]*Facultad de Ingenierías, Universidad Tecnológica de Pereira, Pereira, RI 66003, Colombia*
[†]*Equal contribution first authors*
[\*] *wang1173@purdue.edu, jjv@purdue.edu*

**Abstract:** Soliton formation in on-chip micro-comb generation balances cavity dispersion and nonlinearity and allows coherent, low-noise comb operation. We study the intracavity waveform of an on-chip microcavity soliton in a silicon nitride microresonator configured with a drop port. Whereas combs measured at the through port are accompanied by a very strong pump line which accounts for >99% of the output power, our experiments reveal that inside the microcavity, most of the power is in the soliton. Time-domain measurements performed at the drop port provide information that directly reflects the intracavity field. Data confirm a train of bright, close to bandwidth-limited pulses, accompanied by a weak continuous wave (CW) background with a small phase shift relative to the comb.

## 1. Introduction

Comb generation in high quality factor (Q) microresonators has attracted increasing attention. During the past decade, micro-combs have been demonstrated in a variety of materials, such as silica [1, 2], high-index silica-glass [3], $CaF_2$ [4], $MgF_2$ [5-7], fused quartz [8], SiC [9], and $Si_3N_4$ [7, 10-18]. Similar to observations in fiber-based cavities, studies have shown the possibility to obtain passively mode-locked combs in anomalous dispersion microresonators, where soliton-pulse formation balances cavity dispersion and nonlinearity [2, 6, 7, 11, 16-18]. However, in previous experiments, the comb is measured at the resonator through port, with a very strong overlapping pump line which must be attenuated prior to characterization. Therefore, the intracavity waveform is not fully characterized because the complex amplitude of the intracavity pump field is not known. To obtain such information, here we perform measurements at a microresonator drop port, for which the strong overlapping pump is absent. Drop-port measurements have previously been applied for characterization of comb generation under normal [13, 15] and anomalous dispersion [3], but have not been reported for anomalous dispersion combs with single-soliton formation. Through observations at the drop port, both the efficiency of the intracavity power transfer and the complex amplitude of the CW background field accompanying the soliton are revealed.

This work results in several new findings. First, the soliton spectrum together with the corresponding CW background field are determined directly in the cavity. The difference in the intracavity power in the pump line compared to that in adjacent comb lines is ~10 dB, much less than the power difference (~40 dB) observed at the through port. Second, the intracavity comb shows efficient (up to 85%) power transfer from the pump to the other comb lines, in strong contrast to the poor overall efficiency (<0.5%) observed at the through port. Last, time-domain characterization performed at the drop port confirms that the intracavity field is a single bright pulse close to the bandwidth-limit, accompanied by a weak CW background (power ~ 0.5% of the soliton peak power) with a small but discernible phase shift. Although the phase shift of the CW component is central to the understanding of soliton mode-locking as a self-organization process [19], it has not been previously measured in anomalous dispersion microcavities operating in the soliton regime.

## 2. Device properties and transmission spectra

Our experiment uses a silicon nitride ring resonator with 2 μm (width) × 800 nm (height) waveguide cross-section and 100 μm ring radius (Fig. 1(a)). The ring resonator has anomalous dispersion of $\beta_2$ ~-61 $ps^2$/km measured by the frequency comb assisted spectroscopy [20], close to the simulated value of -52 $ps^2$/km. The gaps between the ring resonator and the through-port and drop-port waveguides are 500 nm and 1000 nm, respectively. The weaker drop-port coupling minimizes the impact of the drop port on the loaded quality (Q) factor [13]. Figure 1(b) shows the through- and drop-port transmission spectra of a quasi-TE mode at ~1551.25 nm. By fitting both the through- and drop-port spectra, the extracted coupling parameters (power per round trip) for the bus and drop waveguides are $4.6 \times 10^{-4}$ and $1.6 \times 10^{-5}$, respectively, while the intrinsic loss parameter is $1.7 \times 10^{-3}$ corresponding to intrinsic Q around $3 \times 10^6$.

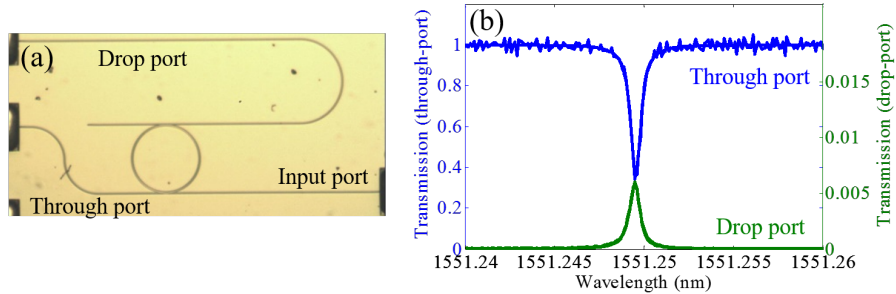

Fig. 1. (a) Image of a silicon nitride microring with through and drop ports. The dust particles are above the upper cladding and will not affect the comb spectrum. (b) Through- (blue trace) and drop-port (green trace) transmission spectra around 1551.25 nm.

The pump laser is amplified to 500 mW (before coupling onto the chip) and scanned across the resonance at a speed of 20 nm/s from high to low frequency. Figure 2 shows the power transmitted at the drop port as a function of time. Due to the thermal nonlinearity, the resonance is distorted into a triangular shape [21]. During the scan a comb begins to form and then enters into a chaotic regime characterized by strong fluctuation in the transmitted power. At time designated zero in our plots, a series of steps down in power are observed, corresponding to formation of solitons with quantized energy, similar to that reported in [2, 6, 7, 18]. Only the single-soliton step (last step before falling out of resonance), which persists over a relatively broad range, can be seen on the time scale plotted in Fig. 2(b). The broad range of the single soliton state at this pumping condition allows us to generate a single soliton almost every time after the soliton kicking process described later. After transitioning to a single soliton, the total cavity power drops to ~16%, of that in chaotic regime. However, as we discuss later, the power balance between pump and comb changes appreciably upon entering the single soliton state. The effective pump detuning is monitored at the through port via the Pound-Drever-Hall (PDH) technique. The transition from noisy operation to stable soliton coincides with the zero crossing point of the PDH signal; the soliton state corresponds to effective red detuning (pump at longer wavelength than resonance) [2, 6].

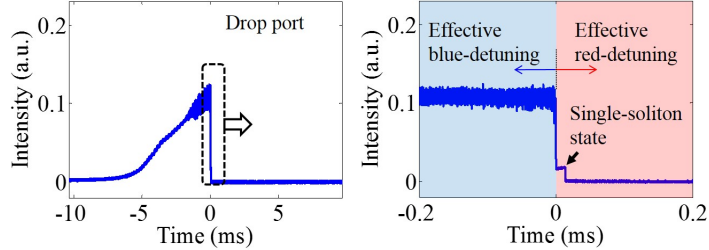

Fig. 2. High power transmission spectrum measured at drop port with resonance around 1551.25 nm. The power step coincides with a transition into effectively red-detuned operation.

## 3. Single-soliton comb spectra and power transfer

Figures 3(a)-(c) show the evolution of comb spectra observed directly at the through-port output, with pumping around 1551.27 nm. The corresponding RF spectra are shown in Fig. 3(d). By red-detuning the pump wavelength into the cavity resonance, the comb spectrum starts with multiple-FSR spacing (Fig. 3(a)) and then evolves into single-FSR spacing (Fig. 3(b)). At this stage the comb exhibits large intensity noise (blue trace in Fig. 3(d)). To overcome thermal instability that hinders realization of stable solitons in the steady state, two different methods are used. The first method is to induce the soliton by a few hundred nsec duration drop in the pump power, as described in [2, 18]. The second method is to control the pump detuning, as explained in [7]. Here the soliton is induced by first forward tuning the laser from blue to red and then stabilized by tuning back by a few picometers from red to blue after the transition from the chaotic state. This process is guided through measurements of the pump laser frequency with a commercial wavemeter with precision down to a few MHz. After the soliton state is reached, the intensity noise drops to the measurement floor, and aside from the strong pump line, a smooth optical spectrum with approximately $sech^2$ shape is observed (Fig. 3(c)). The few small spurs in the spectrum (other than the pump line) are attributed to mode interaction [2, 22]. Figure 3(e) shows another example of single-soliton formation with pumping at 1560.49 nm. The pump line is 39 dB stronger than the adjacent comb spectrum in Fig. 3(c) and 37 dB in Fig. 3(e).

To investigate the intracavity field under soliton formation, we moved the output lensed fiber to the drop port. Figure 4 shows the optical spectra measured at the drop port for the soliton states corresponding to Fig. 3(c) and Fig. 3(e). The most important new observation is that for the drop port spectra, the intensity of the pump line is much closer to that of the adjacent comb lines compared to the through-port spectra. The pump line is 11 dB stronger than the adjacent comb spectrum in Fig. 4(a) and 8 dB in Fig. 4(b), in both cases nearly a 30 dB difference compared to the through port. This suggests that under soliton operation, most of the strong pump power at the through port results from the pump being transmitted directly through the bus waveguide without coupling into the microring. We note that due to the asymmetric coupling, the comb spectrum at the drop port has a similar envelope compared to the through port but is ~15 dB weaker. This difference agrees with the different coupling strengths between the through and drop port (~14.6 dB). Since the amplified spontaneous emission (ASE) between the resonances is filtered out at the drop port, the optical signal-to-noise ratio (OSNR) of the drop-port combs strongly increases in comparison with the through-port combs. This contributes to the ability to perform time-domain measurements of the comb at the drop port, even with low output power. Finally, the peak of the comb envelope is spectrally red-shifted from the pump, which we attribute to the Raman induced soliton shift [16, 17].

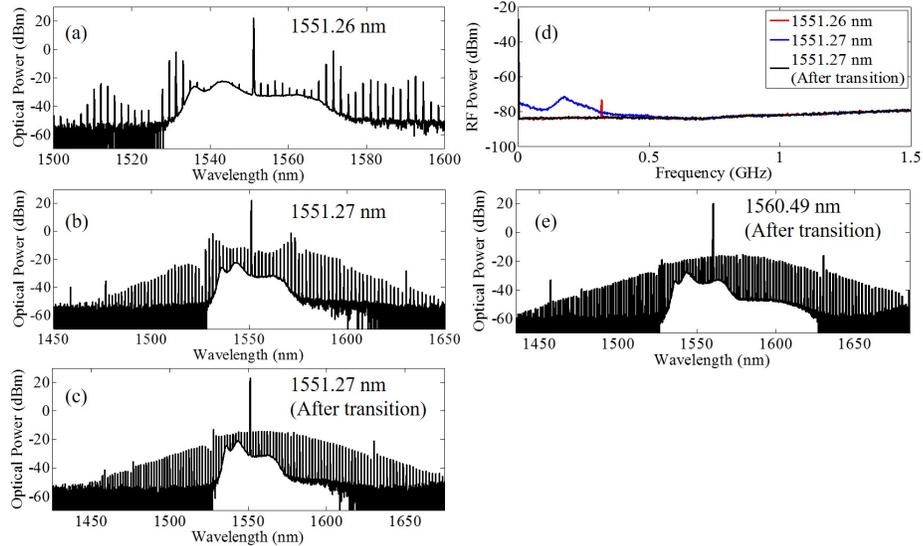

Fig. 3. (a)-(c) Optical spectra in different comb states. (d) RF spectra of the generated combs (100 KHz resolution bandwidth). (e) Another example of single-soliton formation by pumping at 1560.49 nm.

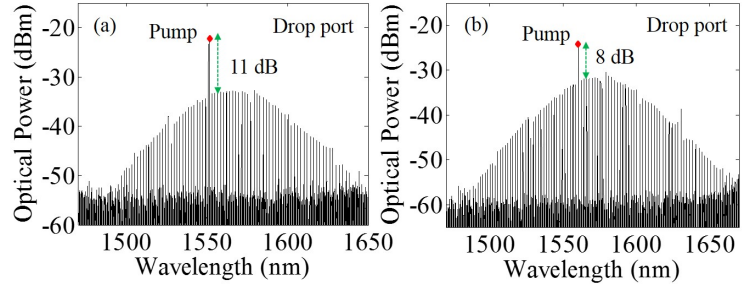

Fig. 4. Drop-port spectra of single-soliton combs pumping at (a) 1551.27 nm and (b) 1560.49 nm.

Table 1 shows the optical power in the output waveguide as well as the corresponding intracavity power for both the chaotic and single-soliton state for the case of 1551.27 nm pumping. The powers in the waveguides are estimated by using the optical spectra and assigning half of the fiber-to-fiber loss (~ 6 dB) to each of the output and input facets. By total comb power we mean the power integrated over all of the comb lines except for the pump line. The intracavity powers in Table 1 are estimated by dividing the drop-port power by the drop-port coupling coefficient.

The very strong pump line observed in all cases at the through port is explained by several factors: (1) the input bus is under-coupled; (2) power transfer into combs aggravates the under-coupling condition [13]; (3) soliton formation is accompanied by a strong effective red-detuning [2, 6, 7] and therefore the power difference is further enhanced after the transition. Due to the large pump power which passes to the through port, the overall efficiency of the generated comb is low (~0.5%). However, we observe, for the first time to our knowledge, that the intracavity field shows efficient (~75% for 1551.27 nm pumping) power transfer from the pump to the comb in the single-soliton regime. Another observation is that upon transitioning from the chaotic to the single-soliton regime, the pump power at the drop port decreases more than ten times, whereas the comb power drops only about three-fold, resulting in a different power balance in the cavity. For 1551.27 nm pumping, the comb accounts for ~46% of the intracavity power in the chaotic state, compared to 75% in the single soliton state. Similar results are obtained for pumping at 1560.49 nm. For the single-soliton state, the drop port power is 6.8 μW in the pump and 38.2 μW in the comb (the intracavity values for pump and comb are 485.7 mW and 2.7 W, respectively). In this case the comb lines account for ~85% of the intracavity power under single soliton operation.

Table 1. Optical Power for Chaotic and Single-soliton State

|  | Pump Power | Total Comb Power |
|---|---|---|
| Chaotic State @ 1551.27 nm |  |  |
| Input Port | 250.0 mW |  |
| Through Port (output) | 206.0 mW | 5.9 mW |
| Drop Port (output) | 124.4 μW | 107.8 μW |
| Intracavity | 7.8 W | 6.6 W |
| Single Soliton @ 1551.27 nm |  |  |
| Input Port | 250.0 mW |  |
| Through Port (output) | 244.3 mW | 1.1 mW |
| Drop Port (output) | 10.8 μW | 30.6 μW |
| Intracavity | 675.0 mW | 1.9 W |

## 4. Time-domain characterization of the single soliton

To investigate the time-domain waveform inside the cavity, we perform intensity autocorrelation measurements at the drop port, based on second harmonic generation in a noncollinear geometry, assisted by a pulse shaper. Measurements performed at the drop port are expected to provide information corresponding directly to the intracavity comb field. In contrast, time-domain measurements which have previously been reported at the through port require in-line filtering (e.g., with a high extinction fiber-Bragg grating) to suppress the strong pump line [2, 6]. Such measurements fail to provide the phase and intensity of the pump line relative to the intracavity comb.

The output from the drop port is relayed to the autocorrelator via standard single-mode fiber through a pair of erbium doped fiber amplifiers (EDFA) and a programmable pulse shaper. We carefully trim the quadratic and the cubic phases applied on the pulse shaper to achieve dispersion compensation for the entire propagation path subsequent to the silicon nitride chip [23]. We test this by injecting a pulse from a mode-locked fiber laser into the fiber link directly after the chip. The link is compensated well enough to measure pulse durations down to a few hundred fsec. The spectrum prior to the autocorrelator is clipped by the pulse shaper passband to the range 1535-1568 nm but is otherwise not intentionally modified, Fig. 5(a).

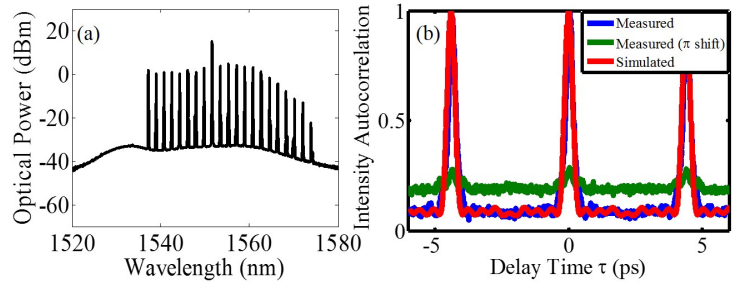

Fig. 5. (a) Optical spectrum after EDFAs and a pulse shaper for 1551.27 nm pumping. (b) Autocorrelation data (blue) and simulation (red) based on measured spectrum assuming flat phase. Autocorrelation data with π phase shift applied to the pump shown in green.

Figure 5(b) shows the measured autocorrelation trace (blue), along with the trace simulated on the basis of the measured power spectrum and assumed flat spectral phase (red). Experimental and simulated traces are in close agreement, suggesting intracavity pulses very close to the bandwidth-limit. Further information is obtained by using the pulse shaper to apply phase shifts to the pump line. With a π phase shift applied to the pump line, the autocorrelation peak is strongly decreased, while the background level of the autocorrelation increases (Fig. 5(b), green). These data demonstrate that the character of the intracavity field depends strongly on the phase of the pump. Furthermore, this phase must be at least relatively close to that of the broadband comb. Such behavior could not be directly verified in previous through-port experiments [2, 6, 7, 16-18] for which the very strong pump line had to be removed or strongly attenuated. Note that the behavior observed contrasts strongly with that reported for mode-locked dark pulses from normal dispersion cavities [15].

To obtain the exact pump phase, we studied the visibility curves, defined as $V(\theta)=(V_0-V_1)/(V_0+V_1)$ [24, 25], where $V_0$ and $V_1$ are the values of the intensity autocorrelation peak and the value half way between the peaks, respectively. $\theta$ is the applied pump phase on the shaper. Figure 6 shows measured visibility data (blue and green dots) vs. phase applied to the pump from two experimental trials. The red trace shows the visibility curve calculated assuming identical intracavity pump and comb phases. The measured visibility is close to that simulated except for a small but discernible phase shift estimated at -0.42 rad. This suggests that the intracavity pump is phase shifted by a complementary amount, i.e., roughly +0.42 rad compared to the comb. A similar small phase offset is obtained in measurements performed for the single soliton with the drop-port spectrum of Fig. 4(b). The sign of the measured phase offset agrees with that predicted from theoretical treatments for anomalous dispersion based on either self-organization in Kerr comb generation [19] or analysis of phase-matching effects [26], assuming that we account for the different sign conventions (in our experiments we use the $e^{j\omega t}$ convention prevalent in ultrafast optics [27], whereas the theoretical treatments use the $e^{-j\omega t}$ convention customary for the Lugiato–Lefever (LL) equation [19, 28]; for further discussion see [15].) We do note that the magnitude of the observed phase shifts are smaller than that expected from the theory. Possible explanations may involve phase shifts associated with mode interactions and/or the Raman induced soliton frequency shift or small propagation effects in the external fiber remaining even after our careful dispersion compensation efforts.

With knowledge of the intracavity pump field, we can predict the intensity profile of the intracavity comb, including background, under the assumption that the soliton is bandwidth-limited throughout its spectrum. Using the full spectrum of Fig. 4(a), we compute a bright pulse with ~74 fs duration (intensity FWHM), accompanied by a weak background (intensity only ~0.5% compared to soliton peak intensity).

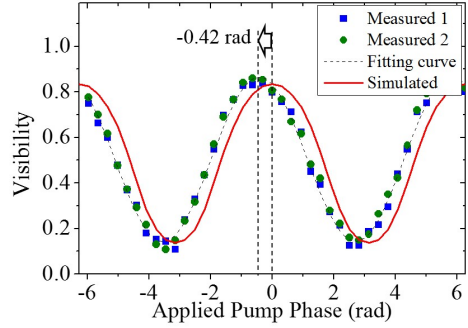

Fig. 6. Simulated (red line) and measured (blue and green dots) visibility of the autocorrelation traces for the comb in Fig. 5(a). The fitting curve of the measured data is shown with a dashed line.

## 5. Conclusion

In conclusion, we have characterized an optical frequency comb operating in the stable single-soliton regime in a silicon nitride microresonator configured with a drop port. Unlike typical through-port experiments in which a very strong overlapping pump field is present, here we are able to provide further information about the comb spectrum, power transfer, and time-domain waveform as it exists in the cavity. Whereas the pump accounts for more than 99% of the power at the through port, we observe that the comb can account for more than 85% of the power at the drop port and in the cavity. An important practical implication is that the typical poor power conversion into the comb may not be fundamental to single soliton operation but may result largely from ineffective coupling. This finding motivates further efforts to optimize coupling under comb generation conditions, perhaps along the lines of early studies such as [29, 30]. Another possibility may be to recycle the large pump power that passes by the resonator with minimal loss for other purposes, which may be attractive for future applications such as coherent telecommunications.

## Acknowledgment

This work was supported in part by the Air Force Office of Scientific Research (AFOSR) (FA9550-15-1-0211), by the DARPA PULSE program (W31P40-13-1-0018) from AMRDEC, and by the National Science Foundation (NSF) (ECCS-1509578).